\documentclass[aps,pra,superscriptaddress,showpacs]{revtex4-1}

\usepackage{epsfig,amsmath}
\usepackage{subfigure}
\usepackage{graphicx}
\usepackage{dcolumn}
\usepackage{stmaryrd}
\usepackage{mathrsfs}
\usepackage{pifont}
\usepackage{amsthm}
\usepackage{amssymb}
\usepackage{bm}
\usepackage{latexsym}
\usepackage[colorlinks=true,linkcolor=blue,citecolor=blue]{hyperref}
\usepackage{color}
\usepackage{bbold}
\usepackage{subfigure}

\newcommand{\lam}{\lambda}
\newcommand{\De}{\Delta}

\newcommand{\ra}{\rangle}

\newcommand{\ga}{\gamma}
\newcommand{\Ga}{\Gamma}
\newcommand{\om}{\omega}

\newcommand{\si}{\sigma}

\newcommand{\vare}{\varepsilon}

\newcommand{\non}{\nonumber}

\begin{document}

\title{Steady bipartite coherence induced by non-equilibrium environment}

\author{Yong Huangfu}
\affiliation{Institute of Atomic and Molecular Physics, Jilin University, Changchun 130012, Jilin, China}

\author{Jun Jing}
\thanks{Author to whom any correspondence should be addressed. Email address: jingjun@zju.edu.cn}
\affiliation{Department of Physics, Zhejiang University, Hangzhou 310027, Zhejiang, China}

\date{\today}

\begin{abstract}
We study the steady state of two coupled two-level atoms interacting with a non-equilibrium environment that consists of two heat baths at different temperatures. Specifically, we analyze four cases with respect to the configuration about the interactions between atoms and heat baths. Using secular approximation, the conventional master equation usually neglects steady-state coherence, even when the system is coupled with a non-equilibrium environment. When employing the master equation with no secular approximation, we find that the system coherence in our model, denoted by the off-diagonal terms in the reduced density matrix spanned by the eigenvectors of the system Hamiltonian, would survive after a long-time decoherence evolution. The absolute value of residual coherence in the system relies on different configurations of interaction channels between the system and the heat baths. We find that a large steady quantum coherence term can be achieved when the two atoms are resonant. The absolute value of quantum coherence decreases in the presence of additional atom-bath interaction channels. Our work sheds new light on the mechanism of steady-state coherence in microscopic quantum systems in non-equilibrium environments. \\

\textbf{Keywords}: steady-state quantum coherence, non-equilibrium environment, master equation, secular approximation
\end{abstract}

\pacs{03.65.Yz, 05.30.-d, 05.70.Ln, 42.25.Kb}

\maketitle

\section{Introduction}

Quantum coherence, including quantum entanglement which is caused by nonlocal coherence, plays an indispensable role in quantum physics. Quantum coherence has been widely applied in many applications, such as quantum heat engines~\cite{Scully_Single Heat Bath_2003,Linke_Single Heat Bath_2003,Scully_noise-induced coherence_2011}, quantum privacy~\cite{Schumacher_Quantum Privacy_1998}, quantum teleportation~\cite{Nielsen_Quantum Computation_2000} and quantum photocells~\cite{Scully_Quantum Photocell_2010}. However, microscopic systems are inevitably in contact with the external environment in all practical applications. It is well-founded that quantum coherence will quickly decay with time in the open-quantum-system scenario, which is called the decoherence process~\cite{Breuer_The Theory of Open Quantum Systems_2002,Maximilian_Decoherence_2005}, which connects the quantum and classical worlds.  Long-lasting (even steady-state) quantum coherence is the subject of active research.

Many protocols to suppress decoherence have been proposed, ranging from external control~\cite{Du_quantum-state-tomography_2016,Yin2016} via bang-bang pulses~\cite{Viola_suppression of decoherence_1998,Viola_Dynamical decoupling_1999,Viola_Universal control_1999}, creating decoherence-free subspaces~\cite{Lidar_Decoherence-Free Subspaces_1998,Kwiat_Decoherence-Free Subspaces_2000} to coherent control by an external field~\cite{Altafini_Coherent control_2004,Jing_with no control_2016,Jing_Overview_2015,Dong_control_2015}. These methods are often performed in the presence of an equilibrium environment with constant temperature and a nearly static state. However, a more practical situation for an open quantum system is the non-equilibrium environment with a non-vanishing temperature difference. The effects of non-equilibrium environments have been studied within various physical systems, for example, the molecule systems~\cite{Zhang_molecular systems_2014} and the electron spin qubits~\cite{Zhang_spin arrays_2017}. Moreover it had been reported that the non-equilibrium environment is closely tied to the quantum phase transition phenomenon~\cite{Takei_superconductor_2008,Oka_superconducting_2010,Fei_coherence_2017}, photoelectric converters~\cite{Su_Photoelectric converters_2016}, and quantum effects in organisms (including photosynthetic light harvesting~\cite{Sarovar_light-harvesting complexes_2010,Lambert_Quantum biology_2013,Fassioli_Photosynthetic light harvesting_2013,Croce_ Natural strategies_2014}, avian magnetoreception~\cite{Lambert_Quantum biology_2013,Hiscock_compass_2016}, decoherence process~\cite{Tegmark_brain processes_2000}, and even quantum cognition in the brain~\cite{Fisher_Quantum cognition_2015}).

In a thermal-equilibrium environment, the diagonal terms (population) in the steady state of the system density matrix will be arranged based on the Boltzmann distribution and the off-diagonal terms (coherence) will be completely removed. The decoherence process to arrive at the steady state can be modeled by the master equation, which is the most popular approach to describing open quantum system~\cite{Breuer_The Theory of Open Quantum Systems_2002,Lindblad_Semigroups_1976}. To derive the master equation in the weak-coupling regime, it is reasonable and effective to employ the secular approximation (equivalent to the rotating wave approximation) in a thermal-equilibrium environment~\cite{Breuer_The Theory of Open Quantum Systems_2002,Allen_Optical Resonance_1987,Benatti_ a common bath_2010,Santos_ a common bath_2014,Barranco_ a common bath_2017}. With this approximation, a previous study~\cite{liao_quantum_2011} using a non-equilibrium environment also showed that the coherence terms of the system steady state vanish when using models of a thermal-equilibrium environment. However, we find the coherence terms in the steady state will not vanish when using the non-secular approximation.

Focusing on non-equilibrium heat environments, we study the quantum coherence of two mutually coupled two-level atomic systems in the steady state. We begin our analysis with a case in which each atom interacts with a separate bath. In this configuration, the heat flux finds a unique path to move from a high-temperature bath to a low-temperature bath. After a sufficiently long evolution, the system of two atoms reaches a steady state because of contact with the heat baths. It was found that the coherence terms~\cite{Li_Steady quantum_2014} are closely related to the non-vanishing flux because of the existence of two baths with a temperature difference, which is inconsistent with the secular approximation. Thus, one method to confirm the validity of the secular approximation is to derive the master equation for the system coupled with a non-equilibrium environment. We find that the absolute value of the coherence term increases with the temperature difference between the two heat baths and then asymptotically tends toward a stable value. With a fixed temperature difference, however, the coherence will find an optimal value with increasing temperature of the cold bath. These two phenomena are most pronounced when the two atoms are resonant.

Next we gradually open more interaction channels between the two resonant atoms and the two baths to study the steady-state coherence. As expected, non-vanishing coherence in the steady state is also found by employing the master equation with no secular approximation. However, the absolute value of steady-state coherence becomes smaller than that in the first case. In the last case, in which each atom interacts with both baths simultaneously, the coherence terms completely vanish. This result suggests that the steady quantum coherence is also sensitive to the configuration of the system-heat-bath interactions. More channels of heat flow lead to reduction of the quantumness of central system.

The remainder of this paper is organized as follows. We introduce our microscopic system consisting of a pair of coupled two-level atoms, and our non-equilibrium environment consisting of two individual heat baths in Sec.~\ref{system}. In Sec.~\ref{caseA}, we analyze the first case in which each atom is coupled with an individual bath, and derive the master equations with and without the secular approximation as general methods for all cases. We compare the two methods and find that the non-secular approximation is useful for modeling the non-vanishing steady-state coherence. In Sec.~\ref{caseBCD}, we consider the other three cases of different configurations and analyze the steady quantum coherence using the non-secular approximation. Discussion and conclusion are presented in Sec.~\ref{conclusion}.

\section{MODEL DISPLAY}\label{system}

\begin{figure}[htbp]
\centering
\includegraphics[width=3.2in]{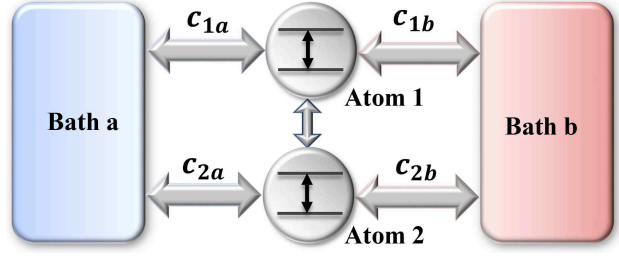}
\caption{The diagram sketch of our model with two coupled atoms interacting with a non-equilibrium environment consisted of a hotter bath-$b$ and a colder bath-$a$. The coupling coefficients $c_{i\alpha}$'s, $i=1,2$, $\alpha=a,b$, determine the configuration of interaction channels.}\label{model}
\end{figure}

As in the diagram in Fig.~\ref{model}, the entire model consists of a system of two two-level atoms $1$ and $2$ and a non-equilibrium environment of two baths $a$ (low-temperature) and $b$ (high-temperature). The bare frequencies of the atoms are $\om_1$ and $\om_2$, respectively. They are coupled with each other by the dipole-dipole interaction. With system-environment interaction, the total Hamiltonian can be divided into three parts: $H=H_{S}+H_{B}+H_{SB}$. Here $H_{S}$ is the Hamiltonian of the central system (hereby we take $\hbar\equiv1$),
\begin{equation}\label{HS1}
H_{S}=\frac{\omega_{1}}{2}\sigma_{1}^{z}+\frac{\omega_{2}}{2}\sigma_{2}^{z}
+\xi(\sigma_{1}^{+}\sigma_{2}^{-}+\sigma_{1}^{-}\sigma_{2}^{+})
\end{equation}
where $\si_{i}^{z}$, $i=1,2$, is the Pauli matrix. The operators $\sigma_{i}^{+}\equiv|e\rangle_{ii}\langle g|$ and $\sigma_{i}^{-}\equiv|g\rangle_{ii}\langle e|$ denote the raising and lowering operators, respectively, where $|g\rangle_{i}$ and $|e\rangle_{i}$ are the ground and excited states of the $i$-th atom, respectively. $\xi$ is the coupling strength between the atoms. The eigen-structure for the system Hamiltonian in Eq.~(\ref{HS1}), which is crucial to obtaining a valid master equation in the following section, is given by
\begin{equation}\label{HS2}
H_S=\sum_{l=1}^{4}\lambda_{l}|\lambda_{l}\rangle\langle\lambda_{l}|
\end{equation}
with displaying eigenvectors and eigenvalues
\begin{alignat}{2}
|\lambda_{1}\rangle & =|ee\rangle, & \quad\lambda_{1} & =\Omega\nonumber\\
|\lambda_{2}\rangle & =\cos\frac{\theta}{2}|eg\rangle+\sin\frac{\theta}{2}|ge\rangle, & \quad\lambda_{2} & =\frac{1}{2}\sqrt{\Delta^{2}+4\xi^{2}}\nonumber \\
|\lambda_{3}\rangle & =-\sin\frac{\theta}{2}|eg\rangle+\cos\frac{\theta}{2}|ge\rangle, & \quad\lambda_{3} & =-\frac{1}{2}\sqrt{\Delta^{2}+4\xi^{2}}\nonumber \\
|\lambda_{4}\rangle & =|gg\rangle, & \quad\lambda_{4}& =-\Omega\nonumber
\end{alignat}
where $\Omega\equiv(\omega_{1}+\omega_{2})/2$, $\Delta\equiv\omega_{1}-\omega_{2}$ and $\cot\theta=\Delta/2\xi$. In this work, the quantum coherence is represented by the off-diagonal terms in the density matrix spanned by $|\lam_i\ra$.

The two heat baths are assumed to be independent of each other. In our model, they can be represented by
\begin{equation}\label{HB}
H_{B}=\sum_{j}\omega_{aj}a_{j}^{\dagger }a_{j}+\sum_{k}\omega_{bk}b_{k}^{\dagger }b_{k}
\end{equation}
where the creation and annihilation operators $a^{\dag}_{j}$ ($b^{\dag}_{k}$) and $a_{j}$ ($b_{k}$) describe the $j$-th ($k$-th) mode in bath-$a$ ($b$) with eigenfrequency $\omega_{aj}$ ($\omega_{bk}$).

The system-bath interaction Hamiltonian is written as
\begin{eqnarray}\label{HSB}
H_{SB}&=&\bigg(c_{1a}\sum_{j}g_ja_{j}\si_{1}^{+}+c_{1b}\sum_{k}g_kb_{k}\si_{1}^{+}+ c_{2a}\sum_{j}g_ja_{j}\si_{2}^{+}+c_{2b}\sum_{k}g_kb_{k}\si_{2}^{+}\bigg)+h.c.
\end{eqnarray}
Here the interaction strengths between atom and bath mode, $g_j$ or $g_k$, are the same, when $j$ and $k$ denote modes with identical frequency. According to the distribution of the coupling strengthes between the $i$-th atom and the $\alpha$-th bath, our model can be analyzed via the following four cases. {\em Case A}: $c_{1a}=c_{2b}=1$ and $c_{1b}=c_{2a}=0$, in which each atom interacts with an individual bath. {\em Case B}: $c_{1a}=c_{2a}=c_{2b}=1$ and $c_{1b}=0$. Here, we add one interaction channel between atom-$2$ and bath-$a$ based on {\em Case A}. Thus, atom-$1$ is only associated with the heat bath at lower temperature whereas atom-$2$ interacts with the two baths simultaneously. {\em Case C}: $c_{1a}=c_{1b}=c_{2b}=1$, and $c_{2a}=0$. Now atom-$2$ is only associated with the heat bath at the higher temperature whereas atom-$1$ interacts with the two baths simultaneously. {\em Case D}: $c_{1a}=c_{1b}=c_{2b}=c_{2a}=1$, in which the two atoms are coupled with both heat baths simultaneously. We will discuss these four cases separately in the following sections.

\section{The coherence in case A}\label{caseA}

Under the assumptions of weak coupling between the microscopic system and the heat baths and an ignorable relaxation time-scale for the heat baths, one can apply the Born-Markov master equation~\cite{Breuer_The Theory of Open Quantum Systems_2002} to investigate the dynamics of the central system in the interaction picture with respect to $H_S+H_B$:
\begin{equation}\label{Markovequation}
\frac{d\rho_{S}(t)}{dt}=-\int_{0}^{\infty}ds\textrm{tr}_{B}[H_{SB}(t),[H_{SB}(t-s),\rho_{S}(t)\otimes\rho_{B}]]
\end{equation}
This equation covers all of the cases that we considered in this paper.

In this section, we focus on {\em Case A}, in which each atom interacts with an individual bath. The interaction Hamiltonian of Eq.~(\ref{HSB}) can then be written as
\begin{equation}\label{CaHSB}
H_{SB}=\left(\sum_{j}g_{j}a_{j}\sigma_{1}^{+}+\sum_{k}g_{k}b_{k}\sigma_{2}^{+}\right)+h.c.
\end{equation}
These two bosonic baths are identical and separable; they remain in thermal equilibrium but at different temperatures. The temperature difference is found to be crucial to non-vanishing steady-state quantum coherence.

In the following sections, we compare the results based on the master equation of Eq.~(\ref{Markovequation}) for the case with the secular approximation and the case without the secular approximation.

\subsection{Master equation with secular approximation }\label{method1}

With the secular approximation, the master equation (in the Schr\"{o}dinger picture) can be explicitly obtained as
\begin{eqnarray}\label{masterequation1}
\dot{\rho}_{S}=\mathcal{L}(\rho_{S})&=&\non -i[H_{S},\rho_{S}]+ \Ga_{1}[L_{\tau_{31}}(\rho_{S})+L_{\tau_{42}}(\rho_{S})]
+\Ga_{2}[L_{\tau_{13}}(\rho_{S})+L_{\tau_{24}}(\rho_{S})]\\
&+&\non \Ga_{3}[L_{\tau_{21}}(\rho_{S})+L_{\tau_{43}}(\rho_{S})]
+\Ga_{4}[L_{\tau_{12}}(\rho_{S})+L_{\tau_{34}}(\rho_{S})]\\
&+& 2(\Lambda _{1}\tau_{42}\rho _{S}\tau_{13}+\Lambda _{2}\tau_{24}\rho _{S}\tau_{31}
+ \Lambda _{3}\tau_{21}\rho _{S}\tau_{34}+\Lambda _{4}\tau_{12}\rho _{S}\tau_{43}+h.c.)
\end{eqnarray}
where $L_{X}(\rho_{S})\equiv2X\rho_{S}X^{\dag}-\{X^{\dag}X, \rho_{S}\}$ with $X$ an arbitrary system operator, $\tau_{ij}\equiv|\lambda_{i}\rangle\langle\lambda_{j}|$, and
\begin{eqnarray}
  \Gamma_{1} &=&\non \cos^{2}\frac{\theta}{2}A_{1}(\varepsilon_{1})+\sin^{2}\frac{\theta}{2}
B_{1}(\varepsilon_{1}), \quad
  \Gamma_{2} =\non \cos^{2}\frac{\theta}{2}A_{2}(\varepsilon_{1})+\sin
^{2}\frac{\theta}{2}B_{2}(\varepsilon_{1}) \\
  \Gamma_{3} &=&\non \sin^{2}\frac{\theta}{2}A_{1}(\varepsilon_{2})+\cos
^{2}\frac{\theta}{2}B_{1}(\varepsilon_{2}),\quad
 \Gamma_{4} = \sin^{2}\frac{\theta}{2} A_{2}(\varepsilon_{2})+\cos
^{2}\frac{\theta}{2}B_{2}(\varepsilon_{2}) \\
  \Lambda _{1} &=&\non \cos^{2}\frac{\theta}{2}A_{1}(\varepsilon_{1})-\sin
^{2}\frac{\theta}{2}B_{1}(\varepsilon_{1}), \quad
  \Lambda _{2} = \cos^{2}\frac{\theta}{2}A_{2}(\varepsilon_{1}) -\sin
^{2}\frac{\theta}{2}B_{2}(\varepsilon_{1}) \\
\Lambda _{3} &=& -\sin ^{2}\frac{\theta}{2}A_{1}(\varepsilon_{2})+\cos
^{2}\frac{\theta}{2}B_{1}(\varepsilon_{2}), \quad
 \Lambda _{4} = -\sin^{2}\frac{\theta}{2}A_{2}(\varepsilon_{2})+\cos
^{2}\frac{\theta}{2}B_{2}(\varepsilon_{2})\label{coeff1}
\end{eqnarray}
Here we define $A_{1}(\varepsilon_{i})=\gamma_{a}(\varepsilon_{i})[N_{a}(\varepsilon _{i})+1]$, $A_{2}(\varepsilon_{i})=\gamma_{a}(\varepsilon_{i})N_{a}(\varepsilon_{i})$, $B_{1}(\varepsilon_{i})=\gamma_{b}(\varepsilon_{i})[N_{b}(\varepsilon_{i})+1]$, and $B_{2}(\varepsilon_{i})=\gamma_{b}(\varepsilon_{i})N_{b}(\varepsilon_{i})$, $i=1,2$. The average thermal excitation numbers are defined according to the Boltzmann distribution $N_{j}(\varepsilon _{i})=1/[\exp(\varepsilon _{i}/T_{j})-1]$ (with $\kappa_{B}=1$, $i=1,2$ and $j=a,b$). $\varepsilon_{1}\equiv\lambda_{13}=\lambda_{24}$. $\varepsilon_{2}\equiv\lambda_{12}=\lambda_{34}$, where $\lambda_{ij}\equiv\lambda_{i}-\lambda_{j}$. The rates $\gamma_{a}(\varepsilon_{i})=\pi\varrho_{a}(\varepsilon_{i})g^{2}(\varepsilon _{i})$ and $\gamma_{b}(\varepsilon_{i})=\pi \varrho_{b}(\varepsilon_{i})g^{2}(\varepsilon _{i})$, where $\varrho_{a}(\varepsilon_{i})$ and $\varrho_{b}(\varepsilon _{i})$ are, respectively, the densities of state at energy $\varepsilon_{i}$ for the two heat baths. Here, we assume $\gamma_{a}(\varepsilon_{i})=\gamma_{b}(\varepsilon_{i})=\ga$. The details of the derivation from Eq.~\eqref{Markovequation} to Eq.~\eqref{masterequation1} can be found in Appendix~\ref{Appcasea}. Based on the master equation~(\ref{masterequation1}), the reduced density matrix of the central system takes a diagonal form in the steady state:
\begin{equation}\label{rho1}
  \rho_{S}(t\rightarrow\infty)=\left(
         \begin{array}{cccc}
           \rho_{11} & 0 & 0 & 0 \\
           0 & \rho_{22} & 0 & 0 \\
           0 & 0 & \rho_{33} & 0 \\
           0 & 0 & 0 & \rho_{44} \\
         \end{array}
       \right)
\end{equation}
where $\rho_{11}=\Gamma_{2}\Gamma_{4}/\Gamma$, $\rho_{22}=\Gamma_{2}\Gamma_{3}/\Gamma$, $\rho_{33}=\Gamma_{1}\Gamma_{4}/\Gamma$, and $\rho_{44}=\Gamma_{1}\Gamma_{3}/\Gamma$ with $\Gamma=(\Gamma_{1}+\Gamma_{2})(\Gamma_{3}+\Gamma_{4})$. This result means that when one applies the master equation with the secular approximation, the coherence terms completely disappear in the steady state even under a non-equilibrium heat environment. This result had been reported in a previous study~\cite{liao_quantum_2011}.

\subsection{Master equation with non-secular approximation }\label{method2}

During the conventional derivation process for the master equation~(\ref{masterequation1}) with secular approximation, the fast evolution terms with factors $e^{\pm i\vare_{12}t}$ ($\varepsilon_{12}\equiv\varepsilon_{1}-\varepsilon_{2}$) are always disregarded. In this subsection, we further analyze these terms in the master equation. With no secular approximation, we add (See Appendix~\ref{Appcasea}) an extra term $R(t)$ to Eq.~(\ref{masterequation1}):
\begin{eqnarray}
\dot{\rho}_{S}&=& \mathcal{L}(\rho_{S})+R(t)
\label{masterequation2}
\end{eqnarray}
where the non-secular term $R(t)$ is defined as 
\begin{eqnarray}
R(t)&=&\non\bigg[\Delta_{1}\big(\tau_{32}\rho_{S}+\tau_{31}\rho_{S}\tau_{12}-\tau_{42}\rho_{S}\tau_{34}\big)
+\Delta_{2}\big(\tau_{12}\rho_{S}\tau_{31}-\rho_{S}\tau_{32}-\tau_{34}\rho_{S}\tau_{42}\big)
\\&+&\non\Delta_{3}\big(\tau_{23}\rho_{S}+\tau_{21}\rho_{S}\tau_{13}-\tau_{43}\rho_{S}\tau_{24}\big)
+\Delta_{4}\big(\tau_{13}\rho_{S}\tau_{21}-\rho_{S}\tau_{23}-\tau_{24}\rho_{S}\tau_{43}\big) \\&+&\non \Delta_{5}\big(\tau_{42}\rho_{S}\tau_{12}-\tau_{31}\rho_{S}\tau_{34}\big) +\Delta_{6}\big(\tau_{12}\rho_{S}\tau_{42}-\tau_{34}\rho_{S}\tau_{31}\big)
\\&+&\non \Delta_{7}\big(\tau_{21}\rho_{S}\tau_{24}-\tau_{43}\rho_{S}\tau_{13}\big) +\Delta_{8}\big(\tau_{24}\rho_{S}\tau_{21}-\tau_{13}\rho_{S}\tau_{43}\big)\bigg]+h.c.
\end{eqnarray}
with
\begin{alignat}{2}
\Delta_{1}&=&\non\sin\frac{\theta}{2}\cos\frac{\theta}{2}[A_{1}(\varepsilon_{1})-B_{1}(\varepsilon_{1})],
\quad\Delta_{2}=\sin\frac{\theta}{2}\cos\frac{\theta}{2}[A_{2}(\varepsilon_{1})-B_{2}(\varepsilon_{1})]\non\\
\Delta_{3}&=&\non\sin\frac{\theta}{2}\cos\frac{\theta}{2}[A_{1}(\varepsilon_{2})-B_{1}(\varepsilon_{2})],
\quad\Delta_{4}=\sin\frac{\theta}{2}\cos\frac{\theta}{2}[A_{2}(\varepsilon_{2})-B_{2}(\varepsilon_{2})]\non\\
\Delta_{5}&=&\non\sin\frac{\theta}{2}\cos\frac{\theta}{2}[A_{1}(\varepsilon_{1})+B_{1}(\varepsilon_{1})],
\quad\Delta_{6}=\sin\frac{\theta}{2}\cos\frac{\theta}{2}[A_{2}(\varepsilon_{1})+B_{2}(\varepsilon_{1})]\non\\
\Delta_{7}&=&\non\sin\frac{\theta}{2}\cos\frac{\theta}{2}[A_{1}(\varepsilon_{2})+B_{1}(\varepsilon_{2})],
\quad\Delta_{8}=\non\sin\frac{\theta}{2}\cos\frac{\theta}{2}[A_{2}(\varepsilon_{2})+B_{2}(\varepsilon_{2})]
\end{alignat}

\begin{figure}[htbp]
\centering
\includegraphics[width=3.4in]{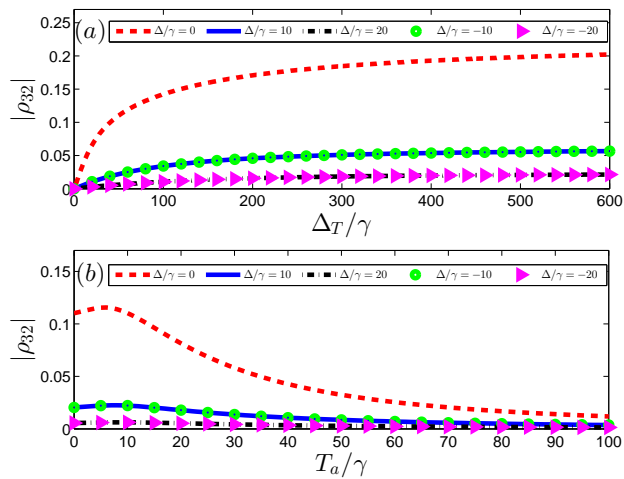}
\caption{The absolute value of $\rho_{32}$ in configuration of {\em Case A} vs. (a) the temperature difference $\Delta_{T}/\ga$ between the two baths and (b) the temperature of the colder bath-$a$ $T_{a}/\ga$, under various detuning between the two atoms $\De/\ga$. We fix $T_{a}/\ga=10$ in (a) and $\Delta_{T}/\ga=50$ in (b). The average frequency of atoms and the coupling strength between atoms are set as $\Omega/\ga=30$ and $\xi/\ga=2$, respectively.  }\label{caseAcom}
\end{figure}

\begin{figure}[htbp]
\centering
\includegraphics[width=3.4in]{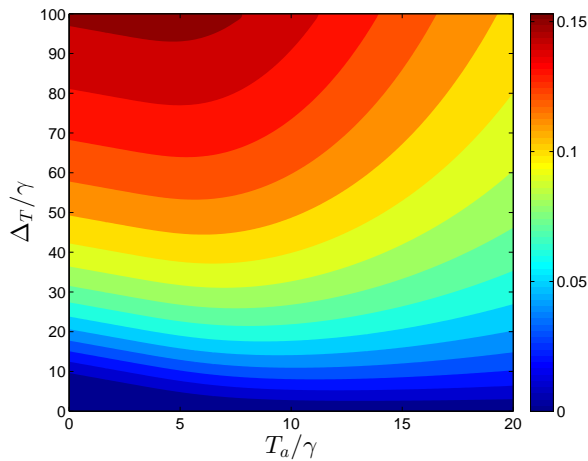}
\caption{The phase diagram of $|\rho_{32}|$ for {\em Case A} in space of the temperature of bath-$a$ $T_{a}/\ga$ and the temperature difference $\Delta_{T}/\ga$. Here we consider the resonance case for the two atoms that $\Delta=0$. The average frequency of atoms and the coupling strength between atoms are set as $\Omega/\ga=30$ and $\xi/\ga=2$, respectively.}\label{caseAtem}
\end{figure}

The steady state based on the master equation~(\ref{masterequation2}) takes the following form
\begin{equation}\label{rho2}
  \rho_{S}(t\rightarrow\infty)=\left(
         \begin{array}{cccc}
           \rho_{11} & 0 & 0 & 0 \\
           0 & \rho_{22} & \rho_{23} & 0 \\
           0 & \rho_{32} & \rho_{33} & 0 \\
           0 & 0 & 0 & \rho_{44} \\
         \end{array}
       \right)
\end{equation}
which is significantly different from Eq.~(\ref{rho1}) under the secular approximation because of the non-vanishing coherent terms. Here, the non-vanishing diagonal and off-diagonal elements are determined by $\textbf{M}X(t)=0$, with
\begin{equation}\non
  \textbf{M}=\left(
               \begin{array}{cccccc}
                 -2\Ga_{13} & 2\Ga_{4}&2\Ga_{2}&0&\Delta_{24}&\Delta_{24}\\
                 2\Ga_{3}&-2\Ga_{14}&0&2\Ga_{2}&\overline{\Delta}_{32}&\overline{\Delta}_{32}\\
                 2\Ga_{1}&0&-2\Ga_{23}&2\Ga_{4}&\overline{\Delta}_{14}&\overline{\Delta}_{14}\\
                 0 & 2\Ga_{1}&2\Ga_{3}&-2\Ga_{24}&-\Delta_{13}&-\Delta_{13}\\
                 \Delta_{13} & \overline{\Delta}_{14} & \overline{\Delta}_{32} & -\Delta_{24}&\mu & 0 \\
                 \Delta_{13} & \overline{\Delta}_{14} & \overline{\Delta}_{32} & -\Delta_{24}&0 & \mu^{*} \\
               \end{array}
             \right)
\end{equation}
where $X(t)=[\rho_{11}\,\, \rho_{22}\,\, \rho_{33}\,\, \rho_{44}\,\, \rho_{32}\,\, \rho_{23}]^{T}$,  $\Ga_{ij}\equiv\Ga_{i}+\Ga_{j}$, $\Delta_{ij}\equiv\Delta_{i}+\Delta_{j}$, $\overline{\Delta}_{ij}\equiv\Delta_{i}-\Delta_{j}$ and $\mu\equiv i\varepsilon_{12}-\Ga_{12}-\Ga_{34}$. As shown by matrix $\textbf{M}$, the off-diagonal (coherence) terms $\rho_{32}$ and $\rho_{23}$ are closely associated with the diagonal terms. In Figs.~\ref{caseAcom} and~\ref{caseAtem}, we plot the absolute value of $\rho_{32}$ using numerical evaluation.

With no loss of generality, in Fig.~\ref{caseAcom}, we assume that $T_b$ is always greater than $T_a$; the temperature difference is defined as $\Delta_{T}=T_{b}-T_{a}$. We set the atom-bath interaction strength $\ga$ as the unit of energy. We fix the value of $T_a$ in Fig.~\ref{caseAcom}(a). The non-vanishing coherence term in our model derives from the temperature difference between the two heat baths. The absolute value of the coherence term $\rho_{32}$ increases monotonically with $\Delta_T$ and then tends toward a stable value. From the five curves denoted by $\De/\ga$ ranging from $-20$ to $20$, $|\rho_{32}|$ is found to increase with decreasing absolute value of the atomic detuning. The curves corresponding to the same absolute value of detuning between the atoms are completely overlapped. In Fig.~\ref{caseAcom}(b), we evaluate the coherence term while increasing the lower temperature $T_a$ and fixing the temperature difference $\Delta_{T}$. We still obtain the greatest coherence $|\rho_{32}|$ in the resonant situation. Note that $|\rho_{32}|$ increases initially and then decreases with increasing $T_a$, which means that there is an optimal point to obtain a significant residue coherence in terms of $T_a$. This optimal point is numerically found to be around $T_a/\ga=8$. The existence of an optimal point is a compromise between two tendencies in physics. On one hand, when $T_a/\ga$ starts from a very low temperature, the higher temperature $T_b$ continues to grow because the temperature difference is fixed. The coherence term will then increase with $T_a$. On the other hand, the ratio of $\Delta_T/T_a$ decreases with increasing $T_a$. The entire non-equilibrium environment, which consists of two individual baths, then tends toward a thermal equilibrium state with the same temperature. This effect degrades the residue coherence.

To obtain a holistic picture of effects of the temperature on the absolute value of the coherence term, we plot a phase diagram in Fig.~\ref{caseAtem} for the resonant situation. It is found that $|\rho_{32}|$ always vanishes when $\Delta_T=0$, i.e., when the non-equilibrium environment reduces to an equilibrium environment. $|\rho_{32}|$ always increased with $\Delta_T$ at a fixed value of $T_a$. However, under a fixed temperature difference $\Delta_T$, to obtain an even larger value of $|\rho_{32}|$, the chosen lower temperature $T_a$ should have a moderate value instead of a smaller one, as we know intuitively. For example, when $\Delta_T/\ga=70$, the chosen value of $T_a/\ga$ should be roughly $6$ to obtain the largest absolute value of the steady-state coherence.

\section{The coherence in the rest cases}\label{caseBCD}

Based on the previous section, we can directly employ the approach of the master equation with non-secular approximation to evaluate the other three cases in our model. Here we provide the resulting master equation for {\em Case B} in Appendix~\ref{Appcaseb}. The derivation processes of the master equations for the other cases are quite similar to {\em Case A} and {\em Case B}.

{\em Case B} is equivalent to adding one atom-bath interaction channel to the model in {\em Case A}. The heat flux now finds more than one paths through the atomic system connecting the high-temperature bath-$b$ to the low-temperature bath-$a$ in this case. The corresponding interaction Hamiltonian of Eq.~(\ref{HSB}) is modified to 
\begin{equation} \label{CbHSB}
H_{SB}=\left[\sum_{j}g_{j}a_{j}(\si_{1}^{+}+\si_{2}^{+})+\sum_{k}g_{k}b_{k}\si_{2}^{+}\right]+h.c.
\end{equation}

With the approach of master equation under the non-secular approximation, we can obtain the steady state in the same form as Eq.~(\ref{rho2}). Here, we concentrate on the coherence term $\rho_{32}$ in the case of two resonant atoms.
\begin{figure}[htbp]
\centering
\includegraphics[width=3.4in]{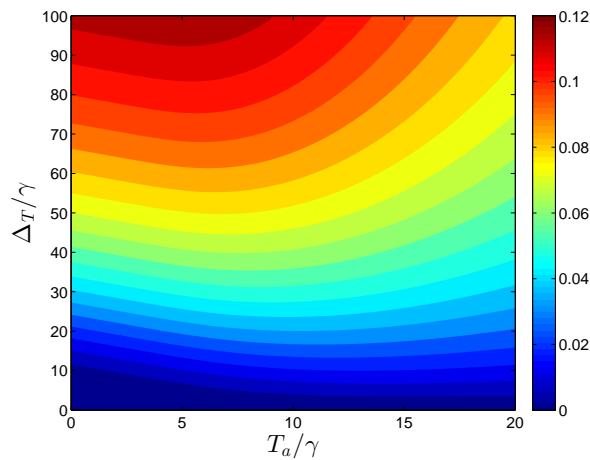}
\caption{The phase diagram of $|\rho_{32}|$ for {\em Case B} in space of the temperature of bath-$a$ $T_{a}/\ga$ and the temperature difference $\Delta_{T}/\ga$. We consider the resonant condition with $\Delta=0$ and other parameters are set as the same as Fig.~\ref{caseAtem}.}\label{caseBtem}
\end{figure}

In the phase diagram of Fig.~\ref{caseBtem}, we find that the overall pattern of $|\rho_{32}|$ does not changed qualitatively, as a function of both the low temperature of $T_{a}$ and the temperature difference $\Delta_{T}$. In the same parameter space as the temperatures, we find that the maximum absolute value of coherence reduces to approximately $0.12$, which is slightly lower than the value in Fig.~\ref{caseAtem}(of approximately $0.15$). This result implies that the modified system-environment interaction configuration in the model has an obvious influence on the steady-state coherence.

\begin{figure}[htbp]
\centering
\includegraphics[width=3.4in]{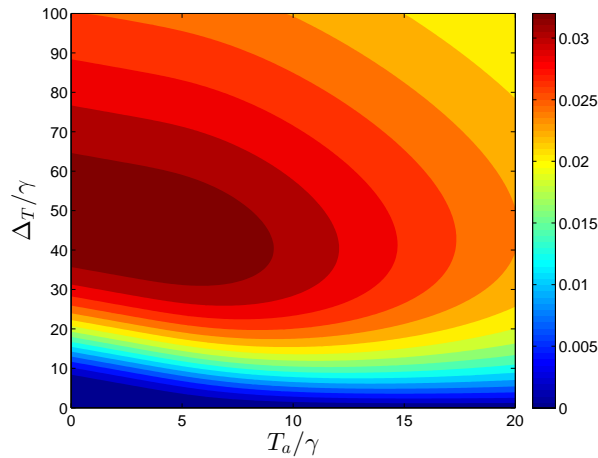}
\caption{The phase diagram of $|\rho_{32}|$ for {\em Case C} in space of the temperature of bath-$a$ $T_{a}/\ga$ and the temperature difference $\Delta_{T}/\ga$. We consider the resonant condition with $\Delta=0$ and other parameters are set as the same as Fig.~\ref{caseAtem}.}\label{caseCtem}
\end{figure}

{\em Case C} is also equivalent to adding one atom-bath interaction channel to the model in {\em Case A}. Now, the heat flux has two paths connecting the system atoms to high-temperature bath-$b$. The interaction Hamiltonian of Eq.~(\ref{HSB}) is modified to
\begin{equation} \label{CcHSB}\non
H_{SB}=\left[\sum_{j}g _{j}a_{j}\si_{1}^{+}+\sum_{k}g_{k}b_{k}(\si_{1}^{+}+\si_{2}^{+})\right]+h.c.
\end{equation}
Following the same procedure as for {\em Case A} and {\em Case B}, we numerically evaluate the residue coherence in the steady state of the central atoms. The temperature-dependent coherence term $|\rho_{32}|$ is shown in Fig.~\ref{caseCtem}. One commonality of these three cases is that a non-vanishing coherence can emerge when $\Delta_{T}>0$. However, the maximum value $|\rho_{32}|$ is approximately one order of magnitude lower than that in {\em Case A} and {\em Case B}. A more interesting result is that $|\rho_{32}|$ no longer increases monotonically with temperature difference at a (properly) fixed value of $T_a$, and then, asymptotically approaches a steady value [(see Fig.~\ref{caseAcom}(a)]. From numerical evaluation, we find that the maximum absolute value of $|\rho_{32}|$ is in the region around $\Delta_{T}/\ga=48$ and $T_{a}/\ga=2$. Beyond this point, increasing the temperature difference gives rise to decreasing coherence. The results of Case B and Case C reflect that the residual coherence will be significantly modified by the effective connection formation between the atoms and heat baths. Particularly in Case C (see Fig.~\ref{caseCtem}), the connection of both atoms with the hotter bath makes the overall model similar to an open system model in a thermal equilibrium bath with a comparatively high temperature. The residual coherence is known to be more fragile in a hotter bath than that in a colder bath. This result explains the difference between Figs.~\ref{caseBtem} and \ref{caseCtem}.

{\em Case D}: we now consider the case in which each atom is simultaneously coupled to both heat baths with a temperature difference. The interaction Hamiltonian is written as
\begin{eqnarray} \label{CdHSB}
H_{SB} &=&\non \left[\sum_{j}g_{j}a_{j}(\si_{1}^{+}+\si_{2}^{+})+\sum_{k}g_{k}b_{k}(\si_{1}^{+}+\si_{2}^{+})\right]+h.c.
\end{eqnarray}
In this case, we also investigate the relationship between the coherence term in the steady state and the temperatures of the two baths. We find that the residual coherence terms completely vanish. Thus the temperature difference between the heat baths is not a necessary condition but a sufficient condition for preserving a non-vanishing coherence in microscopic quantum systems. In this special system-bath interaction configuration, there are various paths for the heat flux to choose in flowing from the high-temperature bath to the low-temperature bath. In {\em Case D}, which has more than two paths, the central quantum system becomes a classical system even in a non-equilibrium environment in the long-time limit, i.e., the steady-state coherence is eventually lost.

\section{Discussion and Conclusion}\label{conclusion}

In the previous sections, we mentioned the relationship between the residual coherence and the heat flux of an atomic system. In general situations, it is well-known that the flux across atom-$1$ consists of two parts, the flux flowing between atoms $J_{1-2}$ and the flux flowing between atom-1 and baths $J_{1-B}$. The heat flux between the two atoms is defined as~\cite{Li_Steady quantum_2014}
\begin{eqnarray}\label{flux}
   J_{1-2}&=& -i\langle[\si^{z}_{1},H_{S}]\rangle =-4\xi \Im \mathfrak{m}(\rho_{32})
\end{eqnarray}
Based on the above Eq.~(\ref{flux}), we conclude that the flux between the two atoms is not only associated with the coherence term $\rho_{32}$ but also the coupling strength between the two atoms. It is beneficial to clarify the physical nature of the steady quantum coherence in a non-equilibrium environment. However, the Markovian master equation applied in this manuscript restrains one to explicitly discussing $J_{1-B}$ for the ``nearly-static'' assumption rather than the environmental state.

In this work, we investigated the residual coherence in the steady-state of two coupled two-level atomic systems under the influence of a non-equilibrium environment consisting of two heat baths at different temperatures. We investigate four configurations in terms of system-heat-bath connections using a master equation with no secular approximation. In certain situations, we can find a significant non-vanishing coherence in the steady state. This study further confirms that the secular approximation is {\em not} appropriate evaluating the steady state of a central system in a non-equilibrium environment. The original high-frequency terms omitted by the secular approximation are crucial to the residual coherence.

In the case in which the two atoms are coupled with individual baths in our model, we find that the absolute value of the coherence increases with increasing temperature difference between the two baths for a fixed temperature of the colder bath. This temperature-driving effect on coherence can be optimized by setting the two atoms at resonance (in bare-frequency) and selecting a suitable value of the colder temperature. By increasing the number of channels connecting the two baths through the atomic system, we gradually decrease the absolute value of the coherence, which will completely vanish when the two atoms are coupled with the two baths simultaneously. A previous study~\cite{Li_Steady quantum_2014} suggested that the quantum steady-state coherence is closely related to the non-equilibrium heat flow. However, our results not only are consistent with the previous result that non-vanishing heat flux supports non-vanishing coherence, but also shows that when the heat flux between the two heat baths has more available paths, it will reduce rather than maintain the residual coherence. Our work is helpful for understanding the steady-state coherence and the quantum-classical transition of a microscopic system in a non-equilibrium environment.

\section*{Acknowledgements}
We acknowledge the grant support from the National Science Foundation of China Grant No. 11575071.

\appendix\label{sec:methods}

\section{Derivation of the master equation for Case A}\label{Appcasea}

We here provide the details of deducing master equation for the system-bath interaction configuration of {\em Case A}, in which each atom solely interacts with an individual heat bath [see the interaction Hamiltonian in Eq.~(\ref{CaHSB})]. Rewrite Eq.~(\ref{CaHSB}) using the eigen-structure in Eq.~(\ref{HS2}), and then transform it into the interaction picture with respect to $H_S+H_B$, we have
\begin{eqnarray}
    H_{SB}(t)&=&  \tau_{13}e^{i\varepsilon_{1}t}B_{13}(t)+\tau_{24}e^{i\varepsilon_{1}t}B_{24}(t)
    +\tau_{12}e^{i\varepsilon_{2}t}B_{12}(t)+\tau_{34}e^{i\varepsilon_{2}t}B_{34}(t)+h.c.
    \label{AppHSB}
\end{eqnarray}
where
\begin{eqnarray*}
B_{12}(t)&=&\sin(\theta/2)A(t)+\cos(\theta/2)D(t),\quad
B_{13}(t)=\cos(\theta/2)A(t)-\sin(\theta/2)D(t)\\
B_{24}(t)&=&\cos(\theta/2)A(t)+\sin(\theta/2)D(t),\quad
B_{34}(t)=-\sin(\theta/2)A(t)+\cos(\theta/2)D(t)
\end{eqnarray*}
with $A(t)\equiv\sum_{j}g_{j}a_{j}e^{-i\omega _{aj}t}$ and $D(t)\equiv\sum_{k}g_{k}b_{k}e^{-i\omega_{bk}t}$.

Substituting Eq.~(\ref{AppHSB}) into the quantum Markov master equation in Eq.~(\ref{Markovequation}), one can get
\begin{widetext}
\begin{eqnarray}\label{Appcaseamq}
\dot{\rho}_{S}&=&\non\int_{0}^{\infty }ds\bigg\{\sum_{(i,j)}\left[\tau _{ij}\rho _{S}\tau
_{ji} e^{i\lambda_{ij}s
}\langle B_{ij}^{\dag }(t-s) B_{ij}(t)\rangle-\tau _{jj}\rho_{S}
e^{-i\lambda_{ij}s}\langle B_{ij}^{\dag }(t)B_{ij}(t-s)\rangle\right.\\
&+&\left.\tau _{ji}\rho _{S}\tau _{ij}e^{-i\lambda_{ij}s}\langle
B_{ij}(t-s)B^{\dag}_{ij}(t)\rangle-\tau _{ii}\rho_{S}
e^{i\lambda_{ij}s }\langle B_{ij}(t)B^{\dag }_{ij}(t-s)\rangle\right]\non\\
&+&\non\sum_{(ij,kl)}\left[\tau _{ij}\rho_{S}\tau _{kl} e^{i\lambda_{ij}s }\langle B_{lk}^{\dag
}(t-s) B_{ij}(t)\rangle+\tau _{lk}\rho_{S}\tau_{ji} e^{i\lambda_{ij}s
}\langle B_{ij}^{\dag }(t-s) B_{lk}(t)\rangle\right]\\
&+&\non \sum_{(mn,pq)}\left[\big(\tau_{mn}\rho_{S}\tau_{pq}\langle B_{qp}^{\dag}(t-s)B_{mn}(t)\rangle+\tau_{pq}\rho_{S}\tau_{mn}\langle B_{mn}(t)B_{qp}^{\dag}(t-s)\rangle\big)e^{i(\lambda_{mn}-\lambda_{qp})t}e^{i\lambda_{qp}s}\right. \\
   &+&\non\left. \big(\tau_{qp}\rho_{S}\tau_{nm}\langle B_{mn}^{\dag}(t-s)B_{qp}(t)\rangle+\tau_{nm}\rho_{S}\tau_{qp}\langle B_{qp}(t)B_{mn}^{\dag}(t-s)\rangle\big)e^{i(\lambda_{qp}-\lambda_{mn})t}e^{i\lambda_{mn}s}\right] \\
   &-&\non \tau_{23}\rho_{S}\langle B_{24}(t)B_{34}^{\dag}(t-s)\rangle e^{i\varepsilon_{12}t}e^{i\varepsilon_{2}s}-\tau_{32}\rho_{S}\langle B_{34}(t)B_{24}^{\dag}(t-s)\rangle e^{-i\varepsilon_{12}t}e^{i\varepsilon_{1}s}\\
   &-&\rho_{S}\tau_{32}\langle B_{13}^{\dag}(t-s)B_{12}(t)\rangle e^{-i\varepsilon_{12}t}e^{i\varepsilon_{1}s}-\rho_{S}\tau_{23}\langle B_{12}^{\dag}(t-s)B_{13}(t)\rangle e^{i\varepsilon_{12}t}e^{i\varepsilon_{2}s}\bigg\}+h.c.
\end{eqnarray}
\end{widetext}
in which the summation parameters $(i,j)=(1,2),(1,3),(2,4)$ and $(2,4)$, $(ij,kl)=(12,43),(13,42),(31,24)$ and $(43,12)$, $(mn,pq)=(13,21),(13,43),(24,21)$ and $(24,43)$. Upon the conventional treatment with secular approximation, Eq.~\eqref{Appcaseamq} is reduced to the terms in the first three line with their hermitian conjugate.

With the definitions of the following damping coefficients as the Fourier transform of correlation functions in Eq.~\eqref{Appcaseamq},
\begin{widetext}
\begin{eqnarray*}
&&\Gamma_{1}=\textrm{Re}\left[\int_{0}^{\infty }ds
\langle B_{24}(t)B^{\dag}_{24}(t-s)\rangle e^{i\varepsilon _{1}s}\right]=
\textrm{Re}\left[\int_{0}^{\infty }ds\langle B_{13}(t)B^{\dag}_{13}(t-s)\rangle e^{i\varepsilon_{1}s}\right]\\
&&\Gamma_{2}=\textrm{Re}\left[\int_{0}^{\infty }ds\langle B^{\dag}_{24}(t-s)B_{24}(t)\rangle e^{i\varepsilon_{1}s
}\right]=\textrm{Re}\left[\int_{0}^{\infty}ds\langle B^{\dag}_{13}(t-s)B_{13}(t)\rangle e^{i\varepsilon _{1}s}\right]\\
&&\Gamma_{3}=\textrm{Re}\left[\int_{0}^{\infty }ds\langle B_{12}(t)B^{\dag}_{12}(t-s)\rangle e^{i\varepsilon _{2}s}\right]=\textrm{Re}\left[\int_{0}^{\infty }ds\langle B_{34}(t)B^{\dag}_{34}(t-s)\rangle e^{i\varepsilon _{2}s
}\right]\\ &&\Gamma_{4}=\textrm{Re}\left[\int_{0}^{\infty }ds\langle B_{12}^{\dag}(t-s)B_{12}(t)\rangle e^{i\varepsilon _{2}s}\right]=\textrm{Re}\left[\int_{0}^{\infty }ds\langle B_{34}^{\dag }(t-s)B_{34}(t)\rangle e^{i\varepsilon _{2}s}\right]
\end{eqnarray*}
\end{widetext}
and
\begin{eqnarray*}
&&\Lambda_{1}=\textrm{Re}\left[\int_{0}^{\infty }ds\langle B_{13}(t)B^{\dag}_{24}(t-s)\rangle e^{i\varepsilon_{1}s}\right],\quad \Lambda_{2}=\textrm{Re}\left[\int_{0}^{\infty }ds\langle B^{\dag}_{13}(t-s)B_{24}(t)\rangle e^{i\varepsilon _{1}s}\right]\\ &&\Lambda_{3}=\textrm{Re}\left[\int_{0}^{\infty }ds\langle B_{34}(t)B^{\dag}_{12}(t-s)\rangle e^{i\varepsilon _{2}s}\right],\quad \Lambda_{4}=\textrm{Re}\left[\int_{0}^{\infty }ds\langle B^{\dag}_{12}(t-s)B_{34}(t)\rangle e^{i\varepsilon_{2}s}\right]
\end{eqnarray*}
we can obtain the master equation~(\ref{masterequation1}) with secular approximation on rotating back to the Schr\"odinger picture. We can further deduce these coefficients in the quantum microscopic model. For example,
\begin{eqnarray}
    \Ga_{1}&\equiv&\non\textrm{Re}\left[\int_{0}^{\infty }ds \langle B_{24}(t)B^{\dag}_{24}(t-s)\rangle e^{i\varepsilon_{1}s}\right]\\
    &=&\non\cos^{2}\frac{\theta}{2}\textrm{Re}\left[\int_{0}^{\infty }ds\sum_{j}g_{j}^{2}\langle a_{j}a^{\dag}_{j}\rangle e^{i(\om_{aj}-\varepsilon_{1})s}\right]+\sin^{2}\frac{\theta}{2}\textrm{Re}\left[\int_{0}^{\infty }ds\sum_{k}g_{k}^{2}\langle b_{k}b^{\dag}_{k}\rangle e^{i(\om_{bk}-\varepsilon_{1})s}\right] \\
   &=&\non\cos^{2}\frac{\theta}{2}\ga_{a}(\varepsilon_{1})
   [N_{a}(\varepsilon_{1})+1]+\sin^{2}\frac{\theta}{2}\ga_{b}(\varepsilon_{1})[N_{b}(\varepsilon_{1})+1]\\
   &=&\non \cos^{2}\frac{\theta}{2}A_{1}(\varepsilon_{1})+\sin^{2}\frac{\theta}{2}B_{1}(\varepsilon_{1})
\end{eqnarray}
where the rates $\gamma_{a}(\varepsilon_{i})=\pi\varrho_{a}(\varepsilon_{i})g^{2}(\varepsilon _{i})$ and $\gamma_{b}(\varepsilon_{i})=\pi \varrho_{b}(\varepsilon_{i})g^{2}(\varepsilon _{i})$. The rest coefficients in master equation~(\ref{masterequation1}) and Eq.~(\ref{coeff1}) can be obtained in a similar way.

In the subsection~\ref{method2}, we focus on the master equation with non-secular approximation. We would pick up the terms which has just been ignored, i.e., those terms in the last four lines in Eq.~(\ref{Appcaseamq}) and their hermitian conjugate. Here we need to define more decoherence coefficients as following:
\begin{eqnarray}
  \Delta_{1} &=&\non \textrm{Re}\left[\int_{0}^{\infty }ds \langle B_{12}(t)B^{\dag}_{13}(t-s)\rangle e^{i\varepsilon_{1}s}\right]
   = -\textrm{Re}\left[\int_{0}^{\infty }ds \langle B_{34}(t)B^{\dag}_{24}(t-s)\rangle e^{i\varepsilon_{1}s}\right]\\
  \Delta_{2} &=&\non \textrm{Re}\left[\int_{0}^{\infty }ds \langle B_{13}^{\dag}(t-s)B_{12}(t)\rangle e^{i\varepsilon_{1}s}\right]
   = -\textrm{Re}\left[\int_{0}^{\infty }ds \langle B_{24}^{\dag}(t-s)B_{34}(t)\rangle e^{i\varepsilon_{1}s}\right]\\
  \Delta_{3} &=&\non \textrm{Re}\left[\int_{0}^{\infty }ds \langle B_{13}(t)B^{\dag}_{12}(t-s)\rangle e^{i\varepsilon_{2}s}\right]
   = -\textrm{Re}\left[\int_{0}^{\infty }ds \langle B_{24}(t)B^{\dag}_{34}(t-s)\rangle e^{i\varepsilon_{2}s}\right]\\
    \Delta_{4} &=&\non \textrm{Re}\left[\int_{0}^{\infty }ds \langle B_{12}^{\dag}(t-s)B_{13}(t)\rangle e^{i\varepsilon_{2}s}\right]
   = -\textrm{Re}\left[\int_{0}^{\infty }ds \langle B_{34}^{\dag}(t-s)B_{24}(t)\rangle e^{i\varepsilon_{2}s}\right]\\
  \Delta_{5} &=&\non \textrm{Re}\left[\int_{0}^{\infty }ds \langle B_{12}(t)B^{\dag}_{24}(t-s)\rangle e^{i\varepsilon_{1}s}\right]
   = -\textrm{Re}\left[\int_{0}^{\infty }ds \langle B_{34}(t)B^{\dag}_{13}(t-s)\rangle e^{i\varepsilon_{1}s}\right]\\
  \Delta_{6} &=&\non \textrm{Re}\left[\int_{0}^{\infty }ds \langle B_{24}^{\dag}(t-s)B_{12}(t)\rangle e^{i\varepsilon_{1}s}\right]
   = -\textrm{Re}\left[\int_{0}^{\infty }ds \langle B_{13}^{\dag}(t-s)B_{34}(t)\rangle e^{i\varepsilon_{1}s}\right]\\
  \Delta_{7} &=&\non \textrm{Re}\left[\int_{0}^{\infty }ds \langle B_{24}(t)B^{\dag}_{12}(t-s)\rangle e^{i\varepsilon_{2}s}\right]
   = -\textrm{Re}\left[\int_{0}^{\infty }ds \langle B_{13}(t)B^{\dag}_{34}(t-s)\rangle e^{i\varepsilon_{2}s}\right]  \\ \Delta_{8} &=&\non \textrm{Re}\left[\int_{0}^{\infty }ds \langle B_{12}^{\dag}(t-s)B_{24}(t)\rangle e^{i\varepsilon_{2}s}\right]
    =-\textrm{Re}\left[\int_{0}^{\infty }ds \langle B_{34}^{\dag}(t-s)B_{13}(t)\rangle e^{i\varepsilon_{2}s}\right]
\end{eqnarray}
Then all of those non-secular terms are collected into
\begin{eqnarray}
  \bar{R}(t)&=&\non\bigg[\Delta_{1}e^{-i\varepsilon_{12}t}\big(\tau_{32}\rho_{S}+\tau_{31}\rho_{S}\tau_{12}-
  \tau_{42}\rho_{S}\tau_{34}\big)  +\Delta_{2}e^{-i\varepsilon_{12}t}\big(\tau_{12}\rho_{S}\tau_{31}-\rho_{S}\tau_{32}
  -\tau_{34}\rho_{S}\tau_{42}\big)    \\   &+&\non \Delta_{3}e^{i\varepsilon_{12}t}\big(\tau_{23}\rho_{S}+\tau_{21}\rho_{S}\tau_{13}
  -\tau_{43}\rho_{S}\tau_{24}\big)    +\Delta_{4}e^{i\varepsilon_{12}t}\big(\tau_{13}\rho_{S}\tau_{21}-\rho_{S}\tau_{23}
  -\tau_{24}\rho_{S}\tau_{43}\big) \\&+&\non \Delta_{5}e^{-i\varepsilon_{12}t}\big(\tau_{42}\rho_{S}\tau_{12}-\tau_{31}\rho_{S}\tau_{34}\big)+ \Delta_{6}e^{-i\varepsilon_{12}t}\big(\tau_{12}\rho_{S}\tau_{42}-\tau_{34}\rho_{S}\tau_{31}\big)
   \\&+&\non\Delta_{7}e^{i\varepsilon_{12}t}\big(\tau_{21}\rho_{S}\tau_{24}-\tau_{43}\rho_{S}\tau_{13}\big) +\Delta_{8}e^{i\varepsilon_{12}t}\big(\tau_{24}\rho_{S}\tau_{21}-\tau_{13}\rho_{S}\tau_{43}\big)\bigg]+h.c.
\end{eqnarray}
We can also deduce these coefficients in the quantum microscopic model. For example,
\begin{eqnarray}
    \Delta_{1}&\equiv&\non\textrm{Re}\left[\int_{0}^{\infty }ds \langle B_{12}(t)B^{\dag}_{13}(t-s)\rangle e^{i\varepsilon_{1}s}\right]\\
    &=&\non \sin\frac{\theta}{2}\cos\frac{\theta}{2}\textrm{Re}\bigg[\int_{0}^{\infty }ds(\sum_{j}g_{j}^{2}\langle a_{j}a^{\dag}_{j}\rangle e^{i(\om_{aj}-\varepsilon_{1})s}
    -\sum_{k}g_{k}^{2}\langle b_{k}b^{\dag}_{k}\rangle e^{i(\om_{bk}-\varepsilon_{1})s})\bigg] \\
   &=&\non\sin\frac{\theta}{2}\cos\frac{\theta}{2}\bigg(\ga_{a}(\varepsilon_{1})[N_{a}(\varepsilon_{1})+1]
   -\ga_{b}(\varepsilon_{1})[N_{b}(\varepsilon_{1})+1]\bigg)\\
   &=&\non \sin\frac{\theta}{2}\cos\frac{\theta}{2}[A_{1}(\varepsilon_{1})-B_{1}(\varepsilon_{1})]
 \end{eqnarray}
The rest coefficients $\De_j$, $j=2,3,\cdots,8$ can be obtained in a similar way. Rotating back to the Schr\"odinger picture, one can obtain the non-secular term $R(t)$ in Eq.~(\ref{masterequation2}).

\section{The master equation with non-secular approximation for Case B}\label{Appcaseb}

Rotating the interaction Hamiltonian of Eq.~(\ref{CbHSB}) into the interacting frame, then substituting it into the master equation Eq.~(\ref{Markovequation}), we can have
\begin{eqnarray}
\dot{\rho}_{S}&=&\non -i[H_{S},\rho_{S}]+\Ga'_{1}L_{\tau_{31}}(\rho_{S})+\Ga'_{2}L_{\tau_{13}}(\rho_{S})
+\Ga'_{3}L_{\tau_{21}}(\rho_{S})+\Ga'_{4}L_{\tau_{12}}(\rho_{S})\\
&+&\non\Ga'_{5}L_{\tau_{42}}(\rho_{S})+\Ga'_{6}L_{\tau_{24}}(\rho_{S})
+\Ga'_{7}L_{\tau_{43}}(\rho_{S})+\Ga'_{8}L_{\tau_{34}}(\rho_{S})\\
&+& 2(\Lambda' _{1}\tau_{42}\rho _{S}\tau_{13}+\Lambda' _{2}\tau_{24}\rho _{S}\tau_{31}
 +\Lambda' _{3}\tau_{21}\rho _{S}\tau_{34}+\Lambda' _{4}\tau_{12}\rho _{S}\tau_{43}+h.c.)+R(t)
\label{masterequation3}
\end{eqnarray}
where
\begin{eqnarray}
  \Ga'_{1} &=& \non (\cos\frac{\theta}{2}-\sin\frac{\theta}{2})^{2}A_{1}(\varepsilon_{1})
  +\sin^{2}\frac{\theta}{2}B_{1}(\varepsilon_{1}),\quad
  \Ga'_{2} = \non  (\cos\frac{\theta}{2}-\sin\frac{\theta}{2})^{2}A_{2}(\varepsilon_{1})
  +\sin^{2}\frac{\theta}{2}B_{2}(\varepsilon_{1}) \\
  \Ga'_{3} &=& \non (\cos\frac{\theta}{2}+\sin\frac{\theta}{2})^{2}A_{1}(\varepsilon_{2})
  +\cos^{2}\frac{\theta}{2}B_{1}(\varepsilon_{2}) ,\quad \Ga'_{4} = \non   (\cos\frac{\theta}{2}+\sin\frac{\theta}{2})^{2}A_{2}(\varepsilon_{2})
  +\cos^{2}\frac{\theta}{2}B_{2}(\varepsilon_{2})\\
  \Ga'_{5} &=& \non (\cos\frac{\theta}{2}+\sin\frac{\theta}{2})^{2}A_{1}(\varepsilon_{1})
  +\sin^{2}\frac{\theta}{2}B_{1}(\varepsilon_{1}) ,\quad\Ga_{6}= \non (\cos\frac{\theta}{2}+\sin\frac{\theta}{2})^{2}A_{2}(\varepsilon_{1})
  +\sin^{2}\frac{\theta}{2}B_{2}(\varepsilon_{1}) \\
  \Ga'_{7} &=& \non (\cos\frac{\theta}{2}-\sin\frac{\theta}{2})^{2}A_{1}(\varepsilon_{2})
  +\cos^{2}\frac{\theta}{2}B_{1}(\varepsilon_{2}) ,\quad \Ga'_{8} =  \non (\cos\frac{\theta}{2}-\sin\frac{\theta}{2})^{2}A_{2}(\varepsilon_{2})
  +\cos^{2}\frac{\theta}{2}B_{2}(\varepsilon_{2})
\end{eqnarray}
and
\begin{eqnarray}
  \Lambda'_{1} &=& \non (\cos^{2}\frac{\theta}{2}-\sin^{2}\frac{\theta}{2})A_{1}(\varepsilon_{1})
  -\sin^{2}\frac{\theta}{2}B_{1}(\varepsilon_{1}),\quad \Lambda'_{2} = \non (\cos^{2}\frac{\theta}{2}-\sin^{2}\frac{\theta}{2})A_{2}(\varepsilon_{1})
  -\sin^{2}\frac{\theta}{2}B_{2}(\varepsilon_{1})  \\   \Lambda'_{3}&=&\non
  (\cos^{2}\frac{\theta}{2}-\sin^{2}\frac{\theta}{2})A_{1}(\varepsilon_{2})
  +\cos^{2}\frac{\theta}{2}B_{1}(\varepsilon_{2}),\quad   \Lambda'_{4} = \non (\cos^{2}\frac{\theta}{2}-\sin^{2}\frac{\theta}{2})A_{2}(\varepsilon_{2})
  +\cos^{2}\frac{\theta}{2}B_{2}(\varepsilon_{2})
\end{eqnarray}
The non-secular term $R(t)$ is defined as
\begin{eqnarray}
  R(t)&=&\non\bigg[\Delta'_{1}\tau_{31}\rho_{S}\tau_{12}+\Delta'_{2}\big(\tau_{42}\rho_{S}\tau_{34}
  -\tau_{32}\rho_{S}\big)
  +\Delta'_{3}\tau_{34}\rho_{S}\tau_{42}+\Delta'_{4}\big(\tau_{12}\rho_{S}\tau_{31}-\rho_{S}\tau_{32})
   \\ &+&\non \Delta'_{5}\tau_{21}\rho_{S}\tau_{13}+ \Delta'_{6}\big(\tau_{43}\rho_{S}\tau_{24}-\tau_{23}\rho_{S}\big)
   +\Delta'_{7}\tau_{24}\rho_{S}\tau_{43}+\Delta_{8}\big(\tau_{13}\rho_{S}\tau_{21}
   -\rho_{S}\tau_{23}\big) \\&+&\non \Delta'_{9}\tau_{42}\rho_{S}\tau_{12}+\Delta'_{10}\tau_{31}\rho_{S}\tau_{34} +\Delta'_{11}\tau_{12}\rho_{S}\tau_{42} +\Delta'_{12}\tau_{34}\rho_{S}\tau_{31}
   \\&+&\non \Delta'_{13}\tau_{21}\rho_{S}\tau_{24}+\Delta'_{14}\tau_{43}\rho_{S}\tau_{13} +\Delta'_{15}\tau_{24}\rho_{S}\tau_{21}+\Delta'_{16}\tau_{13}\rho_{S}\tau_{43}\bigg]+h.c.
\end{eqnarray}
where
\begin{alignat}{2}
\Delta'_{1}&=&\non(\cos^{2}\frac{\theta}{2}-\sin^{2}\frac{\theta}{2})A_{1}(\varepsilon_{1})
-\sin\frac{\theta}{2}\cos\frac{\theta}{2}B_{1}(\varepsilon_{1}) ,\quad\Delta'_{2}=(\cos^{2}\frac{\theta}{2}-\sin^{2}\frac{\theta}{2})A_{1}(\varepsilon_{1})
+\sin\frac{\theta}{2}\cos\frac{\theta}{2}B_{1}(\varepsilon_{1})\\
\Delta'_{3}&=&\non(\cos^{2}\frac{\theta}{2}-\sin^{2}\frac{\theta}{2})A_{2}(\varepsilon_{1})
+\sin\frac{\theta}{2}\cos\frac{\theta}{2}B_{2}(\varepsilon_{1})
,\quad\Delta'_{4}=\non(\cos^{2}\frac{\theta}{2}-\sin^{2}\frac{\theta}{2})A_{2}(\varepsilon_{1})
-\sin\frac{\theta}{2}\cos\frac{\theta}{2}B_{2}(\varepsilon_{1})\\
\Delta'_{5}&=&\non(\cos^{2}\frac{\theta}{2}-\sin^{2}\frac{\theta}{2})A_{1}(\varepsilon_{2})
-\sin\frac{\theta}{2}\cos\frac{\theta}{2}B_{1}(\varepsilon_{2})
,\quad\Delta'_{6}=(\cos^{2}\frac{\theta}{2}-\sin^{2}\frac{\theta}{2})A_{1}(\varepsilon_{2})
+\sin\frac{\theta}{2}\cos\frac{\theta}{2}B_{1}(\varepsilon_{2})\\
\Delta'_{7}&=&\non(\cos^{2}\frac{\theta}{2}-\sin^{2}\frac{\theta}{2})A_{2}(\varepsilon_{2})
+\sin\frac{\theta}{2}\cos\frac{\theta}{2}B_{2}(\varepsilon_{2})
,\quad\Delta'_{8}=\non(\cos^{2}\frac{\theta}{2}-\sin^{2}\frac{\theta}{2})A_{2}(\varepsilon_{2})
-\sin\frac{\theta}{2}\cos\frac{\theta}{2}B_{2}(\varepsilon_{2})\\
\Delta'_{9}&=&\non(\cos\frac{\theta}{2}+\sin\frac{\theta}{2})^{2}A_{1}(\varepsilon_{1})
+\sin\frac{\theta}{2}\cos\frac{\theta}{2}B_{1}(\varepsilon_{1})
,\quad\Delta'_{10}=(\cos\frac{\theta}{2}-\sin\frac{\theta}{2})^{2}A_{1}(\varepsilon_{1})
-\sin\frac{\theta}{2}\cos\frac{\theta}{2}B_{1}(\varepsilon_{1})\\
\Delta'_{11}&=&\non(\cos\frac{\theta}{2}+\sin\frac{\theta}{2})^{2}A_{2}(\varepsilon_{1})
+\sin\frac{\theta}{2}\cos\frac{\theta}{2}B_{2}(\varepsilon_{1})
,\quad\Delta'_{12}=(\cos\frac{\theta}{2}-\sin\frac{\theta}{2})^{2}A_{2}(\varepsilon_{1})
-\sin\frac{\theta}{2}\cos\frac{\theta}{2}B_{2}(\varepsilon_{1})\\
\Delta'_{13}&=&\non(\cos\frac{\theta}{2}+\sin\frac{\theta}{2})^{2}A_{1}(\varepsilon_{2})
+\sin\frac{\theta}{2}\cos\frac{\theta}{2}B_{1}(\varepsilon_{2})
,\quad\Delta'_{14}=(\cos\frac{\theta}{2}-\sin\frac{\theta}{2})^{2}A_{1}(\varepsilon_{2})
-\sin\frac{\theta}{2}\cos\frac{\theta}{2}B_{1}(\varepsilon_{2})\\
\Delta'_{15}&=&\non(\cos\frac{\theta}{2}+\sin\frac{\theta}{2})^{2}A_{2}(\varepsilon_{2})
+\sin\frac{\theta}{2}\cos\frac{\theta}{2}B_{2}(\varepsilon_{2})
,\quad\Delta'_{16}=(\cos\frac{\theta}{2}-\sin\frac{\theta}{2})^{2}A_{2}(\varepsilon_{2})
-\sin\frac{\theta}{2}\cos\frac{\theta}{2}B_{2}(\varepsilon_{2})
\end{alignat}
As we can see, the master equation~(\ref{masterequation3}) for {\em Case B} is dramatically different from the master equation~(\ref{masterequation2}) for {\em Case A}. This difference stems from the modification in the system-bath interaction configuration/Hamiltonian.

\end{document}